\documentstyle{pss}
\input epsf.tex
\begin{document}

\title{Line Shape in the Mirage Experiment}

\author{{\sc A. Aligia}\footnote{) Corresponding author; 
Phone: +54 2944 445170, Fax: +54 2944 445299, e-mail:
aligia@cab.cnea.gov.ar}}

\address{Centro At\'{o}mico Bariloche and Instituto Balseiro, Comisi\'{o}n\\
Nacional de Energ\'{\i }a At\'{o}mica, 8400 Bariloche, Argentina \\
}

\submitted{October 3, 2001}
\maketitle

\hspace{9mm} Subject classification: 75.20.Hr; 73.22.-f

\begin{abstract}
Using a many-body theory, we calculate the change in differential
conductance $\Delta dI/dV$ after adding an impurity either on a clean
surface or inside an elliptic quantum corral. Using the same set of
parameteres for both cases, the qualitative features of the voltage
dependence of $\Delta dI/dV$ observed in recent experiments, are reproduced.
\end{abstract}

\section{Introduction}

In recent experiments a Co atom (acting as a magnetic impurity) was placed
at one focus of an elliptic corral, and a depression in the differential
conductance $dI/dV$ as a function of voltage (a signature of the Kondo
effect) was observed not only at this focus, but also at the other one \cite
{man}. The space dependence of $\Delta dI/dV$ ( $dI/dV$ after substracting
the corresponding result for the empty corral) reflects mainly the density
of the eigenstate of two-dimensional free electrons confined in the corral
which lies at the Fermi energy. This space dependence has been reproduced by
several theories. The voltage dependence is more subtle and requires in
principle a many-body calculation of the Kondo resonance. However, most
theories treat the interactions in a phenomenological way. Many-body effects
were included either by perturbation theory \cite{ali} or by numerical
diagonalization in a restricted subspace \cite{hal}. Unfortunately, since the
separation of the relevant eigenstates of the corral (those with a sizeable
hybridization with the impurity) is larger than the Kondo temperature \cite
{ali,por}, these numerical results cannot reproduce the observed line shape.

Here we extend our previous approach \cite{ali} to the clean surface and
take into account the direct tunneling between the tip and the impurity 
\cite{sch}, in order to compare the line shape of $\Delta dI/dV$ for an
impurity inside the corral or on a clean surface.

\section{The model}

The simplest description of the problem includes a non-degenerate highly
correlated impurity state and the surface states near the Fermi energy $%
\epsilon _{F}$. For the clean Cu(111) surface, the latter are uncoupled from
bulk states for wave vectors inside the neck of the Fermi surface \cite{hub}%
. Band structure calculations indicate that these states have a weight
larger than 60\% in the first layer. The remaining weight is mainly due to
other surface states at different energies \cite{eu}. Taking a standard 
$r^{-7/2}$ distance dependence between $sp$ and $d$ electrons, and
considering that the impurity atom enters in the three-fold coordinated
position at the surface, one obtains that the hybridization of the impurity
with the surface layer is nearly 17 times larger than that with the second
layer. Since the density of $sp$ states at the first layer is also larger 
\cite{eu}, this information suggests that the direct hybridization of the
impurity with bulk states can be neglected. Taking into account that the
distance between Cu atoms is smaller than the Fermi wave length $2\pi
/k_{F}\sim $ 30\AA , the hybridization of the impurity with a surface state $%
j$ can be approximated as $V\lambda \varphi _{j}(R_{i})$, where $R_{i}$ is a
two-dimensional vector denoting the position of the impurity, $\varphi
_{j}(r)$ is the wave function of state $j$ normalized as $\int |\varphi
_{j}(r)|^{2}d^{2}r=1$, $\lambda $ is a length scale which we take as the
square root of the surface per Cu atom ($\lambda =2.38$ \AA ), and $V$ is an
energy which would represent the hopping between one atom of the surface and the
impurity in a tight-binding description. For the clean surface, $j$ labels
the allowed wave vectors of extended Bloch states in a large area with
periodic boundary conditions, while for the mirage experiment, the $\varphi
_{j}(r)$ are localized inside the corral.

The above considerations lead to the following Anderson model: 
\begin{eqnarray}
H &=&\sum_{j\sigma }\varepsilon _{j}c_{j\sigma }^{\dagger }c_{j\sigma
}+E_{d}\sum_{\sigma }d_{\sigma }^{\dagger }d_{\sigma }+Ud_{\uparrow
}^{\dagger }d_{\uparrow }d_{\downarrow }^{\dagger }d_{\downarrow }  \nonumber
\\
&&+V\lambda \sum_{j\sigma }[\varphi _{j}(R_{i})d_{\sigma }^{\dagger
}c_{j\sigma }+{\rm H.c.}].  \label{ham}
\end{eqnarray}
Here $c_{j\sigma }^{\dagger }$ and $d_{\sigma }^{\dagger }$ create an
electron on the $j^{th}$ conduction eigenstate and the impurity respectively.

\section{Approximations and relevant equations}

At sufficiently low temperatures, the differential conductance $dI/dV$ when
the probing tip is at position $r$, is proportional to the density of the
mixed state $f_{\sigma }(r)=\lambda \sum_{j}\varphi _{j}(r)c_{j\sigma
}+qd_{\sigma }$. \cite{sch}:

\begin{equation}
\frac{dI(V)}{dV}\sim \rho _{f}(\epsilon _{F}+eV)=-\frac{1}{\pi }{\rm Im}
G_{f}(\epsilon _{F}+eV),  \label{cond}
\end{equation}
where $G_{f}(\omega )=\langle \langle f_{\sigma };f_{\sigma }^{\dagger
}\rangle \rangle _{\omega }$ is the Green function of $f_{\sigma }(r)$ and $%
q $ is the ratio between tunneling matrix elements between impurity and tip
and between impurity and surface. It is different from zero only for $r$
very near $R_{i}$. The change in $dI/dV$ after adding the impurity is
proportional to the corresponding change in $\rho _{f}$. From the equations
of motion:

\begin{equation}
\Delta G_{f}(r,\omega )=(VG_{c}^{0}(r,R_{i},\omega
)+q)(VG_{c}^{0}(R_{i},r,\omega )+q)G_{d}(\omega ),  \label{gf}
\end{equation}
where $G_{d}(\omega )=\langle \langle d_{\sigma };d_{\sigma }^{\dagger
}\rangle \rangle _{\omega }$ is the impurity Green function and $G_{c}^{0}$
is the conduction electron Green function for $V=0$:

\begin{equation}
G_{c}^{0}(r,r^{\prime },\omega )=\sum_{j}\frac{\lambda ^{2}\varphi
_{j}^{*}(r)\varphi _{j}(r^{\prime })}{\omega +i\delta -\varepsilon _{j}},
\label{g0}
\end{equation}
where $\delta $ is a positive infinitesimal. For the clean surface, the
eigenvalues $\varepsilon _{j}$ form a continuum of constant density of
states $\rho _{0}=0.045$ states/(eV site), using an effective mass 0.38 times
the electron mass $m_e$ \cite{eu}. For a hard wall elliptic corral, the 
$\varepsilon _{j}$ are discrete. For a finite confinement potential, the
eigenstates inside the corral become resonances and the spectrum is
continuous again. However, from results borrowed mainly from nuclear
physics, we know that $G_{c}^{0}$ inside the corral can be expressed as a
discrete sum of complex poles \cite{gar}. Thus, the form of $G_{c}^{0}$ for
the hard wall corral is retained, but $\delta $ becomes finite. Here we
retain the values of $\varepsilon _{j}$ and $\varphi _{j}(r)$ of the hard
wall corral calculated before \cite{ali} 
up to an energy of 1 eV above $\epsilon_F$. 
This approximation is justified by
comparison with the Green function which results for a scattering potential
consisting of delta functions regularly spaced at the boundary of the
ellipse \cite{cor}. For all continuum states around a resonance, the
wave function inside the corral is essentially the same, differing only
outside the corral. In adition, with an effective mass 0.378 $m_e$, 
the hard wall results for $\varepsilon _{j}$ around $\epsilon_F$ reproduce
the experimental results. For simplicity, we take $\delta$ independent of $j$.

The impurity Green function is given by:

\begin{equation}
G_{d}^{-1}(\omega )=\omega -\tilde{E}_{d}-V^{2}G_{c}^{0}(R_{i},R_{i},
\omega )-\Sigma (\omega ),  \label{gd}
\end{equation}
where in our approach, $\tilde{E}_{d}$ is a renormalized $d$ level, and 
$\Sigma (\omega )$ is
the self energy calculated in second order in $U$ \cite{ali}. Because of the
limitations of the perturbative approach, we have chosen a moderate value of 
$U=1$ eV, and $\tilde{E}_{d}\simeq \epsilon _{F}$ (near the symmetric case).
We have taken $V=0.64$ eV, to lead to a half width at half maximum near 
$T_{K}=4$ meV for the spectral density $\rho _{d}(\omega )$ of the impurity
on a clean surface \cite{man}. The resulting impurity resonant level width
is $\Delta \simeq 58$ meV. For this value of $U/\Delta $, the position of
the peak is not adequately given by perturbation theory. Then we used
independent values of $\tilde{E}_{d}$ for the impurity on the clean surface
or inside the corral, to control this position. The other parameters of the
model are the same in both situations. The width $\delta $ is the only free
parameter to control the line shape for the impurity inside the corral.

\section{Results}

\begin{figure}
\epsfxsize=16cm
\vbox{\hskip -3cm \epsffile{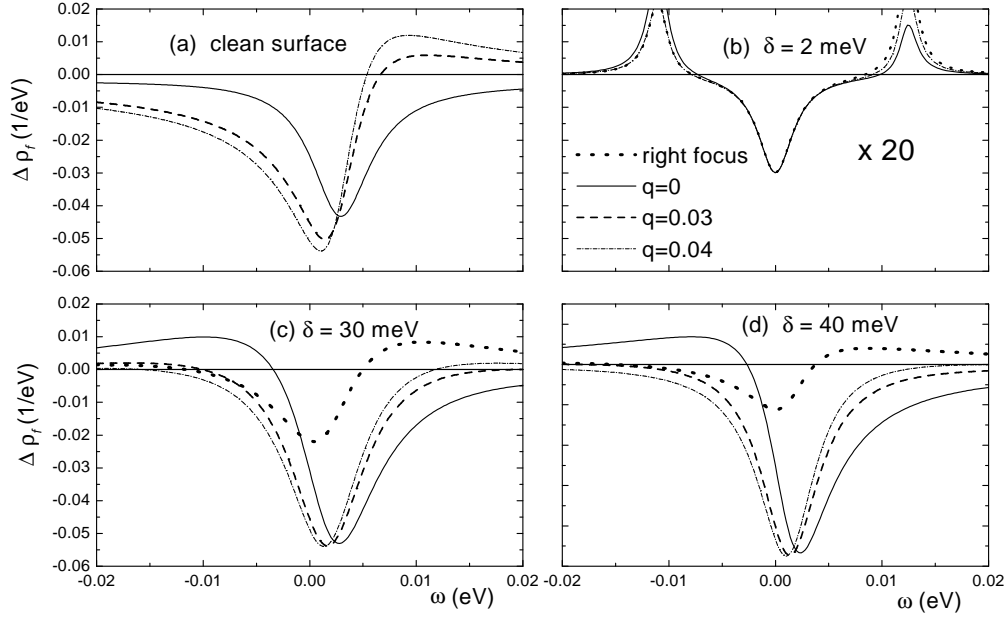}}
\caption{Change of the spectral density of the mixed state $f$ for
(a) clean surface, (b-d) elliptical corral with impurity at the left focus,
and the indicated values of $\delta$ and $q$. The tip is at the impurity 
position except for the dotted lines which correspond to the empty focus
in the corral}
\end{figure}

In Fig. 1 (a), we show $\Delta \rho _{f}$, proportional to $\Delta dI/dV$, at
the impurity position for the clean surface and several values of $q$. The
result for $q=0$ is symmetrical and proportional to $\rho _{d}(\omega )$.
The experimental $\Delta dI/dV$ is larger for positive sample bias,
corresponding to positive energy $\omega $. This asymmetry can be achieved
increasing $q$. However, the experimental line shape shows a relative
maximum also at negative $\omega $, in contrast to our result. This feature
seems difficult to explain. Except for this fact, the observed line shape is
qualitatively reproduced for $q$ between 0.03 and 0.04.

In Fig. 1 (b), (c) and (d)  we show $\Delta \rho _{f}$ for an 
elliptical corral with
eccentricity $e=1/2$, and semimajor axis $a=71.3$ \AA\ \cite{man}, with an
impurity placed at the left focus, for several values of $\delta $ and $q$,
and for both foci. If $\delta <T_{K}$, the Kondo resonance at $\epsilon _{F}$
is absent. Instead, both $\rho _{d}$ and $\rho _{f}$ present two narrow
peaks at both sides of the Fermi energy. In the absence of the impurity, 
$\rho _{f}$ has a peak at $\epsilon _{F}$ which corresponds to the state 42 
($\epsilon _{F}=\varepsilon _{42}$). As a consequence, the difference 
$\Delta \rho _{f}$
present two narrow peaks above zero and a depression with negative $\Delta
\rho _{f}$ (see Fig. 1 (b)). For $\delta \gg T_{K}$ (Fig. 1 (c) and (d)), 
a Kondo resonance quite similar to that
of the clean surface is formed. However, due to the asymmetry of the
amplitude of the wave functions of the corral above and below $\epsilon _{F}$ 
at the impurity site, the line shape is asymmetric for $q=0$, being higher
at negative energies. This asymmetry is opposite to the experimentally
observed in the case of the clean surface. Thus, while a positive value of 
$q$ renders the line shape asymmetric for a clean surface, it
restores a rather symmetric line shape in the case of the corral, as
experimentally observed.

As $\delta $ increases, the magnitude of the depression observed at the
empty focus decreases. This is due to an increasing negative interference of
the corral state 42, which is even under interchange of both foci, with
other odd states, whose weight at $\epsilon _{F}$ increase with $\delta $ 
\cite{ali}. For $\delta \sim 40$ meV, the main aspects of the observed 
space and voltage dependence of  $\Delta dI/dV$ are reproduced. 

\section{Conclusion}

An Anderson model in which the magnetic impurity is hybridized with surface
states, which are extended in the case of a clean surface or resonances in
presence of a quantum corral, is able to reproduce the qualitative features
of the observed change in the voltage dependence of the differential
conductance $\Delta dI/dV$ (in particular its asymmetry) in both cases. This
requires a small direct tunneling between the impurity and probing tip, when
the tip is above the impurity. The rather symmetric line shape observed
above the impurity at the focus in the quantum corral is the result of the
compensation of two effects which alone would produce an asymmetry in
opposite directions: the direct  tunneling between the impurity and tip, and
the particular electronic structure of the resonances inside the corral,
above and below the Fermi energy. Work to study the line shape in other 
physical situations, for example for a mirage out of focus \cite{ali,por,wei} 
is in progress. 

\section{Acknowledgments}

This work benefitted from PICT 03-00121-02153 of ANPCyT and PIP 4952/96 of
CONICET. We are partially supported by CONICET.

\end{document}